\documentclass[page-classic]{epl2} 

\def\be{\begin{eqnarray}}
\def\en{\end{eqnarray}}
\def\non{\nonumber\\}

\def\ra{\rangle}

\def\sl{\!\!\!\slash}
\def\prd{{Phys. Rev. D}~}
\def\prl{{ Phys. Rev. Lett.}~}
\def\plb{{ Phys. Lett. B}~}

\def\epjc{{ Eur. Phys. J. C}~}
\newcommand{\acp}{{\cal A}_{CP}}
\title{Branching Ratio and CP Asymmetry of $B_s \to K^*_0(1430)\pi$ Decays in the PQCD Approach}
\shorttitle{Branching Ratio and CP Asymmetry of $B_s \to K^*_0(1430)\pi$ Decays in the PQCD Approach} 

\author{Zhi-Qing Zhang\inst{}}

\institute{
  \inst{} Department of Physics, Henan University of
Technology,
\\Zhengzhou, Henan 450052, P.R.China}


\abstract{
In the two-quark model
supposition for $K_0^{*}(1430)$,
the branching ratios and the direct CP-violating
asymmetries for decays $\bar B_s^0\to K^{*0}_0(1430)\pi^0, K^{*+}_0(1430)\pi^-$
are studied by employing the perturbative QCD factorization approach. We find that although these two decays are both
tree-dominated, the ratio of their penguin to tree contributions are very different: there is only a few percent for the decay
$\bar B_s^0\to K^{*+}_0(1430)\pi^-$, while about
$37\%$ in scenario I, even $51\%$ in scenario II for the decay $\bar B_s^0\to K^{*0}_0(1430)\pi^0$. It results that
these two decays have very different values in the branching ratios and the direct CP asymmetries.
The branching ratio
of the decay $\bar B_s^0\to K^{*+}_0(1430)\pi^-$ is at the order of $10^{-5}$, and its direct CP asymmetry is about $(20-30)\%$. While for the decay
$\bar B_s^0\to K^{*0}_0(1430)\pi^0$,
its direct CP-violating asymmetry is very large and about $90\%$, but it is difficult to measure it, because the branching ratio for this channel is small and only $10^{-7}$
order.}

\begin{document}

\maketitle

\section{Introduction}
Along with many scalar mesons found in experiments, more and more efforts have been made to study the scalar meson spectrum
theoretically \cite{nato,jaffe,jwei,baru,celenza,stro,close1}. There are two typical schemes for their classification \cite{nato,jaffe}. Scenario I (SI): the nonet mesons below
1 GeV, including $f_0(600), f_0(980), K^*_0(800)$, and $a_0(980)$, are
usually viewed as the lowest lying $q\bar q$ states, while the nonet
ones near 1.5 GeV, including $f_0(1370), f_0(1500)/f_0(1700),
K^*_0(1430)$, and $a_0(1450)$, are suggested as the first excited
states. In scenario II (SII), the nonet mesons near 1.5 GeV are
treated as $q\bar q$ ground states, while the nonet mesons below 1
GeV are exotic states beyond the quark model, such as four-quark
bound states. In order to
uncover the inner structures of these scalar mesons, many factorization approaches are used to research
the $B$ meson decay modes with a final state scalar meson, such as the generalized factorization approach
\cite{GMM}, QCD factorization approach
\cite{CYf0K,ccysp,ccysv}, and perturbative QCD (PQCD) approach
\cite{zqzhang1,zqzhang2,zqzhang3,zqzhang4,zqzhang5}.

Whether $K^{*0}_0(1430)$ belongs to
the first excited state (scenario I) or the lowest lying state (scenario II) is an interesting question, which impacts on its hadronic parameters,
such as form factor, decay constant and Gegenbauer moment. For example, the form factor $F^{BK^*_0}_0(q^2)$ is defined as
\be
F^{BK^*_0}_0(q^2)=\frac{F^{BK^*_0}_0(0)}{1-a(q^2/m^2_B)+b(q^2/m^2_B)^2},
\en
where the parameters $a, b$ are obtained from the fitting procedure among the region $0<q^2<10 GeV^2$.
They have different values in two scenarios by using the covariant light-front quark model \cite{cch}, $a=0.59, b=0.09$
for scenario I and $a=0.44, b=0.05$
for scenario II. From the potential model calculation, we know that the decay constant and the form factor for scenario
I have opposite signs \cite{ccysp}, while it is not the case for scenario II.
From QCD sum rule calculation \cite{dsdu}, one can find different masses for the scalar meson $s\bar q (q=u, d)$
under the different scenario assumptions:$m(s\bar q)=1.410\pm0.049$ GeV for scenario II and $m(s\bar q)>2.0$ GeV for scenario I,
so the authors considered that scenario II is more favored.
$B$ meson decays offer a promising opportunity to investigate this question: Here $K^*_0(1430)$ can be treated as a $q\bar q$
state in both scenarios, it is easy to make quantitative predictions in the two-quark  model
supposition, so we would like to use the PQCD approach to calculate the branching ratios
and the CP-violating
asymmetries for decays $\bar B_s^0\to K^{*0}_0(1430)\pi^0, K^{*+}_0(1430)\pi^-$. Certainly, we use the hadronic parameters
derived from QCD sum-rule method in our calculations. Comparing the theoretical prediction with the future experimental data will
indicate information on the structure of $K^{*0}_0(1430)$.
In the following, $K^*_0(1430)$ is denoted as $K^*_0$ in some places for convenience.

\section{ The perturbative QCD  calculation} \label{results}

Under the two-quark model for the scalar meson $K^*_0(1430)$ supposition,
the amplitudes for decays $\bar B^0_s\to K^*_0\pi$
 can be conceptually written as the convolution,
\be
 {\cal
A}(B_s \to  K^*_0\pi) &\sim &\int\!\! d x_1 d x_2 d x_3 b_1 d b_1 b_2 d
b_2 b_3 d b_3 \non && \cdot \mathrm{Tr} \left [ C(t) \Phi_{B_s}(x_1,b_1)
\Phi_{K^*_0}(x_2, b_2) \Phi_{\pi}(x_3,b_3) H(x_i, b_i, t) S_t(x_i)\,
e^{-S(t)} \right ], \quad \label{eq:a2}
\en
where $b_i(i=1,2,3)$ is the
conjugate space coordinate of $k_{iT}$, and $t$ is the largest
energy scale in function $H(x_i,b_i,t)$. Here
$\mathrm{Tr}$ denotes the trace over Dirac and color indices, $C(t)$
is the Wilson coefficient evaluated at scale $t$, which
includes the hard dynamics being from $m_W$ scale down to $t\sim\mathcal{O}(\sqrt{\bar{\Lambda} M_{B_s}})$.
The function
$H(k_1,k_2,k_3,t)$ describes  the six-quark hard scattering kernel, which consists of the effective four quark operators and a
hard gluon to connect the spectator quark in the decay.
In order to smear the end-point singularity on $x_i$,
the jet function $S_t(x)$ \cite{li02}, which comes from the
resummation of the double logarithms $\ln^2x_i$, is used.
The last term $e^{-S(t)}$ is the Sudakov form factor which suppresses
the soft dynamics effectively \cite{soft}. So this hard part $H$ can be perturbatively calculated.
Here $x_{i}(i=1,2,3)$ is the momenta fraction of the antiquark in each meson.
There are the same conventions with Refs. \cite{zqzhang4,zqzhang5} in our calculations.

In the standard model, the related weak effective
Hamiltonian $H_{eff}$ mediating the $b\to d$ type transitions can be written as
\be
\label{eq:heff} {\cal H}_{eff} = \frac{G_{F}} {\sqrt{2}} \,
\left[\sum_{p=u,c}V_{pb} V_{pd}^* \left (C_1(\mu) O_1^p(\mu) +
C_2(\mu) O_2^p(\mu) \right) -V_{tb} V_{td}^*\sum_{i=3}^{10} C_{i}(\mu) \,O_i(\mu)
\right] .
\en
where the local four-quark operator $Q_i (i=1,...,10)$ and the corresponding Wilson coefficient $C_i$ can be found
in Ref. \cite{buras96}.
$V_{p(t)b}, V_{p(t)d}$ are the CKM matrix elements.

From the leading order Feynman diagrams for each considered channel, it is easy to get the analytic
formulas for the amplitudes corresponding to $(V-A)(V-A)$, $(V-A)(V+A)$ and $(S-P)(S+P)$ operators, which are similar to those of $B\to f_0(980)K(\pi), f_0(1500)K(\pi)$ \cite{zqzhang4,zqzhang5}.
We just need to replace some corresponding wave functions, Wilson coefficients, and parameters.

Combining the contributions from different diagrams, the total decay
amplitudes for these decays can be written as
\be
\sqrt{2}{\overline{\cal M}}(K^{*0}_0\pi^0)&=&\xi_u\left[M_{eK^*_0}C_2
+F_{eK^*_0}a_2\right]-\xi_t\left[F_{eK^*_0}\left(-a_4-\frac{1}{2}(3C_7+C_8)
+\frac{5}{3}C_9+C_{10}\right)\right.\non &&\left.+F^{P2}_{eK^*_0}(a_6-\frac{1}{2}a_8)+M_{eK^*_0}(-\frac{C_3}{3}+\frac{C_9}{2}+\frac{3C_{10}}{2})
-(M^{P1}_{eK^*_0}+M^{P1}_{aK^*_0})
\right.\non &&\left.\times (C_5-\frac{C_7}{2})+M^{P2}_{eK^*_0}\frac{3C_{8}}{2}-M_{aK^*_0}(C_3-\frac{1}{2}C_9)-F_{aK^*_0}(a_4-\frac{1}{2}a_{10})
\right.\non &&\left.-F^{P2}_{aK^*_0}(a_6-\frac{1}{2}a_8)\right],\\
\overline{\cal M}(K^{*+}_0\pi^-)&=&\xi_u\left[M_{eK^*_0}C_1
+F_{eK^*_0}a_1\right]-\xi_t\left[F_{eK^*_0}\left(a_4+a_{10}\right)+F^{P2}_{eK^*_0}\left(a_6+a_{8}\right)\right.\non &&\left.+M_{eK^*_0}(C_3+C_9)
+M^{P1}_{eK^*_0}(C_5+C_7)
+M_{aK^*_0}(C_3-\frac{1}{2}C_9)\right.\non &&\left.+M^{P1}_{aK^*_0}(C_5-\frac{1}{2}C_7)+F_{aK^*_0}(a_4-\frac{1}{2}a_{10})
+F^{P2}_{aK^*_0}(a_6-\frac{1}{2}a_8)\right],
\en
where $F_{e(a)K^*_0}$ and $M_{e(a)K^*_0}$
are the $\pi$ meson emission (annihilation) factorizable
contributions and nonfactorizable contributions from penguin operators respectively. The upper label $T$ denotes the
contributions from the tree operators. $P1$ and $P2$ denote the contributions from the $(V-A)(V+A)$ and $(S-P)(S+P)$ type operators, respectively.
The others are the contributions from the $(V-A)(V-A)$ type ones.
The combinations of the Wilson coefficients are defined as usual:
 \be
a_{1}(\mu)&=&C_2(\mu)+\frac{C_1(\mu)}{3}, \quad
a_2(\mu)=C_1(\mu)+\frac{C_2(\mu)}{3},\non
a_i(\mu)&=&C_i(\mu)+\frac{C_{i+1}(\mu)}{3},\quad
i=3,5,7,9,\non
a_i(\mu)&=&C_i(\mu)+\frac{C_{i-1}(\mu)}{3},\quad
i=4, 6, 8, 10.\label{eq:aai} \en

\section{Numerical results and discussions} \label{numer}
In  the two-quark picture, the
scalar decay constant $\bar {f}_{K^*_0}$ for the scalar meson
$K^*_0$ can  be defined as  \be \langle K^*_0(p)|\bar
q_2q_1|0\ra=m_{K^*_0}\bar {f}_{K^*_0}, \label{fbar} \en where
$m_{K^*_0}(p)$ is the mass (momentum) of $K^*_0$. The light-cone
distribution amplitudes for the  scalar meson $K^*_0$
can be written as \be \langle K^*_0(p)|\bar q_1(z)_l
q_2(0)_j|0\rangle &=&\frac{1}{\sqrt{2N_c}}\int^1_0dx \; e^{ixp\cdot
z}\non && \times \{ p\sl\Phi_{K^*_0}(x)
+m_{K^*_0}\Phi^S_{K^*_0}(x)+m_{K^*_0}(n\sl_+n\sl_--1)\Phi^{T}_{K^*_0}(x)\}_{jl},\quad\quad\label{LCDA}
\en where $n_+$ and $n_-$ are lightlike vectors:
$n_+=(1,0,0_T),n_-=(0,1,0_T)$. The
twist-2 light-cone distribution amplitude $\Phi_{K^*_0}$ can be expanded in the Gegenbauer
polynomials: \be \Phi_{K^*_0}(x,\mu)&=&\frac{\bar
f_{K^*_0}(\mu)}{2\sqrt{2N_c}}6x(1-x)\left[B_0(\mu)+\sum_{m=1}^\infty
B_m(\mu)C^{3/2}_m(2x-1)\right]. \en
As for the twist-3 distribution amplitudes $\Phi_{K^*_0}^S$ and $\Phi_{K^*_0}^T$, we adopt the asymptotic form:
\be
\Phi^S_{K^*_0}&=& \frac{1}{2\sqrt {2N_c}}\bar f_{K^*_0},\,\,\,\,\,\,\,\Phi_{K^*_0}^T=
\frac{1}{2\sqrt {2N_c}}\bar f_{K^*_0}(1-2x).
\en
The decay constant $\bar f_{K^*_0}$ and
the Gegenbauer moments $B_1,B_3$ of distribution amplitudes for
$K^*_0(1430)$ have been calculated in the QCD sum rules\cite{ccysp}, which are listed as
\be
{\rm scenario I:} B_1&=&0.58\pm0.07, B_3=-1.2\pm0.08, \bar f_{K^*_0}=-(300\pm30){\rm MeV},\\
{\rm scenario II:}B_1&=&-0.57\pm0.13, B_3=-0.42\pm0.22, \bar f_{K^*_0}=(445\pm50){\rm MeV},\quad
\en

The other input parameters used in the numerical calculations are specified below \cite{pdg08}
\be
f_{B_s}&=&230 MeV, M_{B_s}=5.37 GeV, M_W=80.41 GeV, \\
\alpha&=&100^\circ\pm20^\circ, \tau_{B_s}=1.470\times 10^{-12} s,\\
|V_{ub}|&=&3.93\times10^{-3}, V_{ud}=0.974, \\
 |V_{td}|&=&8.1\times10^{-3}, V_{tb}=1.0.
\en
It is noticed that there is dramatic difference for the centra value of CKM angle $\alpha$ between the previous and the present PDG values.
The present PDG value of $\alpha$ is $89.0^{+4.4}_{-4.2}$ \cite{pdg10}, which is shown in the following figures.

In the $B_s$-rest frame, the decay width of $\bar B^0_s\to
K^*_0(1430)\pi$ can be written as \be
\Gamma=\frac{G_F^2(1-r^2_{K^*_0})}{32\pi m_{B_s}}|\overline{\cal
M}|^2, \en where $\overline{\cal M}$ is the total decay amplitude of
each considered decay, which has been given in the previous section,
and the mass ratio $r_{K^*_0}=m_{K^*_0}/M_{B_s}$. ${\overline{\cal M}}$ can
be rewritten as \be \overline{\cal M}=
V_{ub}V^*_{ud}T-V_{tb}V^*_{td}P=V_{ub}V^*_{ud}\left[1+ze^{i(-\alpha+\delta)}\right]
\label{ampde}, \en where $\alpha$ is the Cabibbo-Kobayashi-Maskawa
weak phase angle, and $\delta$ is the relative strong phase between
the tree and the penguin amplitudes, which are denoted as "T" and
"P," respectively. The term $z$ describes the ratio of penguin to
tree contributions and is defined as \be
z=\left|\frac{V_{tb}V^*_{td}}{V_{ub}V^*_{ud}}\right|\left|\frac{P}{T}\right|.
\en If one relates $\overline {\cal M}$ with ${\cal M}$, which is the
total decay amplitude for the corresponding conjugated decay mode,
it is easy to rewrite the decay width $\Gamma$ as \be
\Gamma=\frac{G_F^2(1-r^2_{K^*_0})}{32\pi
m_{B_s}}|V_{ub}V^*_{ud}T|^2\left[1+2z\cos\alpha\cos\delta+z^2\right].
\en So the CP-averaged branching ratio for each considered decay is
defined as \be {\cal B}=\Gamma\tau_{B_s}/\hbar,\label{brann} \en
where $\tau_{B_s}$ is the life time of $B_s$ meson.
\begin{largetable}
\begin{center}
\caption{ Decay amplitudes for decays $\bar B^0_s\to  K^{*0}_0(1430)\pi^0, K^{*+}_0(1430)\pi^-$ ($\times 10^{-2} \mbox {GeV}^3$).}
\begin{tabular}{ccccccccc}
\hline \hline  &&$F^T_{eK^*_0}$&$F_{eK^*_0}$ & $M^T_{eK^*_0}$ &$M_{eK^*_0}$& $M_{aK^*_0}$ &$F_{aK^*_0}$\\
\hline
$\bar B^0_s\to K^{*0}_0(1430)\pi^0$ (SI) &&-13.4&-1.2&$-6.4+4.5i$&$0.15-0.08i$&$-0.04+0.08i$ &$-2.7-6.5i$\\
$\bar B^0_s\to K^{*0}_0(1430)\pi^0$ (SII) &&16.5&1.7&$0.2+3.9i$&$0.01-0.09i$&$0.20+0.11i$ &$0.2+8.6i$\\
$\bar B^0_s\to K^{*+}_0(1430)\pi^-$ (SI) &&121.3&0.1&$4.0-2.7i$&$-0.14+0.09i$&$0.05-0.11i$ &$3.8+9.1i$\\
$\bar B^0_s\to K^{*+}_0(1430)\pi^-$ (SII) &&-220.7&0.6&$0.1-2.7i$&$-0.01+0.11i$&$-0.27-0.16i$ &$-0.3-12.2i$\\
\hline \hline
\end{tabular}\label{amp}
\end{center}
\end{largetable}

Using the input parameters and the wave functions as specified in this section, it is easy to get the values of the
factorizable and nonfactorizable amplitudes from the emission and annihilation topology diagrams
of the considered decays in both scenarios,
which are listed in Table 1. It is noticed that each penguin amplitude in the table includes the contributions from three kinds of operators.
The emission factorizable and nonfactorizbale diagram contributions from the tree
operators are dominant.
In the decay $\bar B_s^0\to K^{*+}_0(1430)\pi^-$, the amplitude $F^T_{eK^*_0}$ is enhanced by the large Wilson coefficients $C_2+C_1/3$, which induce the tree operator contributions to
be absolutely dominant and the ratio $P/T$ is only $8\%$ in scenario I, $5.5\%$ in scenario II. Certainly, for the other channel
$\bar B^0_s\to K^{*0}_0(1430)\pi^0$, the Wilson coefficients associated with the amplitude $F^T_{eK^*_0}$ are $C_2/3+C_1$, which are
color suppressed. We know that the sign of $C_2/3$ is positive while the sign of $C_1$ is negative, which can cancel each other mostly.
So the amplitude $F^T_{eK^*_0}$ is highly suppressed compared with the one for the channel $\bar B_s^0\to K^{*+}_0(1430)\pi^-$. So that the influence from the penguin
contributions becomes important in the decay $\bar B^0_s\to K^{*0}_0(1430)\pi^0$.
The penguin operator contributions are mainly from the factorizable emission diagrams and annihilation diagrams, the latter
are more important and provide a large imagine part, which often induces a large direct CP-violating asymmetry (shown in Table 2). From Table 1, one can find that
the ratio $P/T$ for the decay $\bar B^0_s\to K^{*0}_0(1430)\pi^0$ is large and about $37\%$ for scenario I, $51\%$ for scenario II.

Using the amplitudes as specified in Table 1, we can calculate
the branching ratios of the considered modes, which are listed in Table 2. The uncertainties are mainly from the $B_s$ meson shape parameter $\omega_b$, the decay constant $\bar f_{K^*_0}$,
the Gegenbauer moments $B_1$ and $B_3$ of the scalar meson $K^*_0$.
\begin{largetable}
\begin{center}
\caption{ CP-averaged branching ratios and direct CP asymmetries of decays $\bar B^0_s\to  K^{*0}_0(1430)\pi^0, K^{*+}_0(1430)\pi^-$.
The theoretical errors of branching ratios correspond to the uncertainties due to variation of (i) the $B_s$ meson shape parameter $\omega_b$,
(ii) the decay constant $\bar f_{K^*_0}$, (iii) the Gegenbauer moments $B_1$ and $B_3$ of the scalar meson $K^*_0$. While the errors
of direct CP asymmetries are mainly from (i), (iii) and the CKM angle $\alpha=(100\pm20)^\circ$.    }
\begin{tabular}{ccccc}
\hline \hline  &&Branching ratios &Direct CP asymmetries (in \%)\\
\hline
$\bar B^0_s\to K^{*0}_0(1430)\pi^0$ (SI) &&$(4.5^{+1.4+0.9+0.4}_{-1.1-0.9-0.4})\times 10^{-7}$&$93.6^{+4.8+1.0+0.0}_{-7.3-1.0-8.7}$\\
$\bar B^0_s\to K^{*0}_0(1430)\pi^0$ (SII) &&$(4.1^{+1.0+1.0+0.6}_{-0.7-0.1-0.6})\times 10^{-7}$&$95.5^{+1.2+3.4+4.0}_{-8.7-4.1-13.9}$\\
$\bar B^0_s\to K^{*+}_0(1430)\pi^-$ (SI) &&$(1.2^{+0.5+0.2+0.1}_{-0.3-0.2-0.1})\times 10^{-5}$&$28.1^{+4.9+0.7+0.0}_{-4.4-0.7-2.4}$\\
$\bar B^0_s\to K^{*+}_0(1430)\pi^-$ (SII) &&$(3.7^{+1.4+0.9+0.8}_{-1.0-0.8-0.7})\times 10^{-5}$&$21.0^{+3.4+2.1+0.0}_{-3.1-2.5-2.5}$\\
\hline \hline
\end{tabular}\label{amp}
\end{center}
\end{largetable}
\begin{figure}[t,b]
\begin{center}
\includegraphics[scale=0.6]{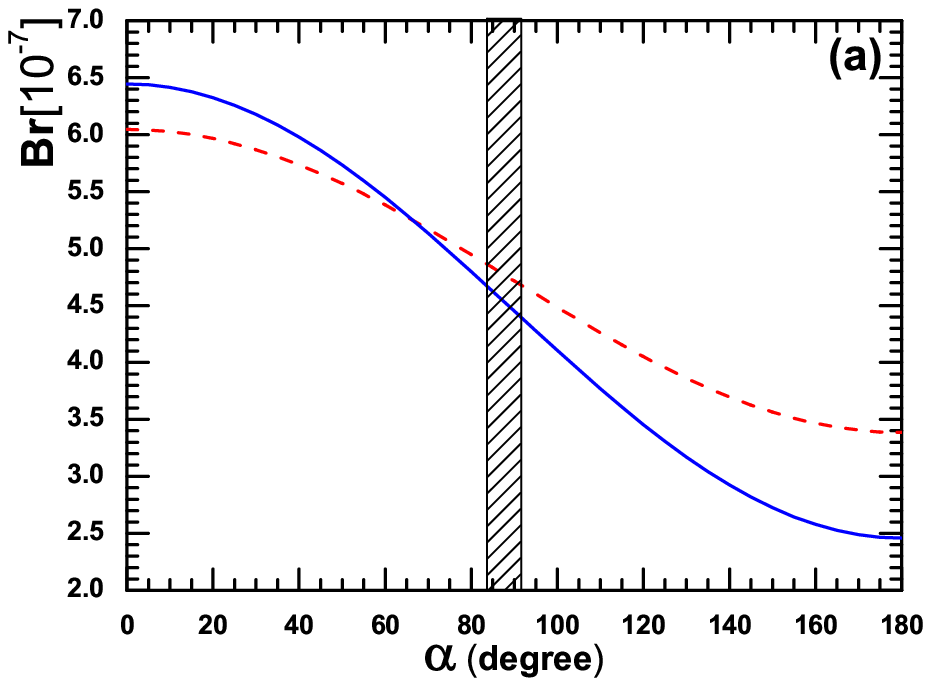}
\includegraphics[scale=0.6]{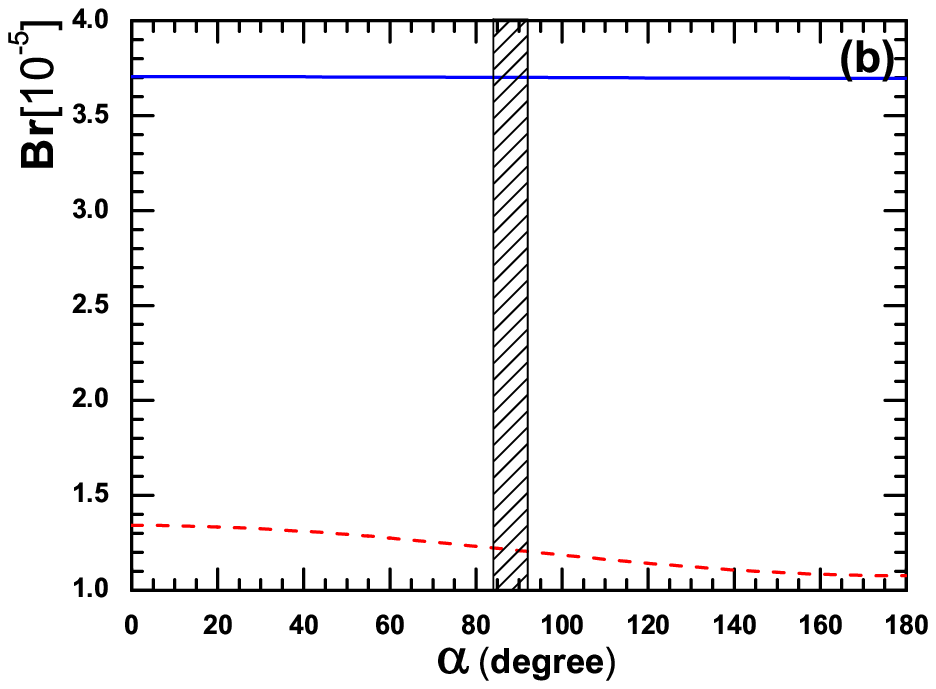}
\vspace{0.3cm} \caption{The dependence of the branching ratios for $\bar B_s^0\to K^{*0}_0(1430)\pi^0$ (a)
and $\bar B_s^0\to K^{*+}_0(1430)\pi^-$ (b)
on the
Cabibbo-Kobayashi-Maskawa angle $\alpha$. The dashed (solid)  curves are plotted in scenario I (II).
The vertical bands show the range of $\alpha$: $89.0^{+4.4}_{-4.2}$ \cite{pdg10}.}\label{fig2}
\end{center}
\end{figure}
From the results, one can find that the branching
ratio of the decay channel $\bar B_s^0\to K^{*+}_0(1430)\pi^-$ is about two order lager than that of $\bar B_s^0\to K^{*0}_0(1430)\pi^0$.
It is because that the former receives a much
larger $\pi^0$ emission factorizable diagram amplitude than the latter. It is interesting to contrast the predictions for decays
$\bar B_s^0\to K_0\pi$ and those for $\bar B_s^0\to K^{*}_0(1430)\pi$, it seems to exist some similar point: the QCD factorization approach predicted the
branching ratio of the
decay $\bar B_s\to K^+_0\pi^-$ was about $(1.02^{+0.59}_{-0.52})\times10^{-5}$ and $(4.9^{+6.3}_{-3.5})\times10^{-7}$ \cite{Beneke} for the decay
$\bar B_s^0\to K^0_0\pi^0$. Certainly, the present experimental result for the decay $\bar B_s\to K^+_0\pi^-$ is about
$(5.0\pm1.1)\times10^{-6}$ \cite{hfag}. If
the future experimental value for the decay $\bar B_s^0\to K^{*+}_0(1430)\pi^-$ also falls into $10^{-6}$ order, which is less than
our prediction, it is might because the following two reasons: First, the value of the transition from factor $\bar B^0_s\to K^*_0(1430)$ predicted
by the PQCD approach is larger than the data. Second, we only calculate in the leading-order. The higher order contributions might give some corrections
to the leading order results. Some contributions from the next-to-leading-order (NLO) corrections have been calculated for the decays $B\to K\pi$
\cite{lihn},
and the results show that their branching ratios decrease by about $20\%$ after including the NLO effects.
These effects might also have an influence on our considered decays.

The dependence of the branching ratios for the decays $\bar B^0_s\to K^{*0}_0(1430)\pi^0$ and $\bar B^0_s\to K^{*+}_0(1430)\pi^-$
on the Cabibbo-Kobayashi-Maskawa angle $\alpha$
is displayed in Fig.1. Compared with Fig.1(a), we know that the branching ratio of the decay $\bar B^0_s\to K^{*+}_0(1430)\pi^-$ is insensitive to
the angle $\alpha$, which is shown in Fig.1(b). From the definition of the CP-averaged branching ratio shown in Eq.(\ref{brann}), one can find that
if the branching
ratio is insensitive to the angle $\alpha$, the coefficient of $\cos\alpha$ must be near zero. For the decay $\bar B^0_s\to K^{*+}_0(1430)\pi^-$, the value of $z$ is very small in
both scenarios, about $0.1\sim0.2$, so $z^2$ is smaller and can be neglected compared with $1$, at the same time, the strong phase angle
$\delta$ is about $89^\circ$ in scenario II, so the value of $(1+2z\cos\alpha\cos\delta+z^2)$ is close to $1$. Although the strong phase angle $\delta$ for
scenario I is
not so large (about $68.6^\circ$), the small value of $z$ makes the branching ratio in this scenario also insensitive to $\alpha$.

Now, we turn to the evaluations of the direct CP-violating asymmetries of
the considered decays in the PQCD approach. The direct CP-violating asymmetry can be defined as
\be
\acp^{dir}=\frac{ |\overline{\cal M}|^2-|{\cal M}|^2 }{
 |{\cal M}|^2+|\overline{\cal M}|^2}=\frac{2z\sin\alpha\sin\delta}
{1+2z\cos\alpha\cos\delta+z^2}\;. \label{dirdifine}
\en
\begin{figure}[t,b]
\begin{center}
\includegraphics[scale=0.6]{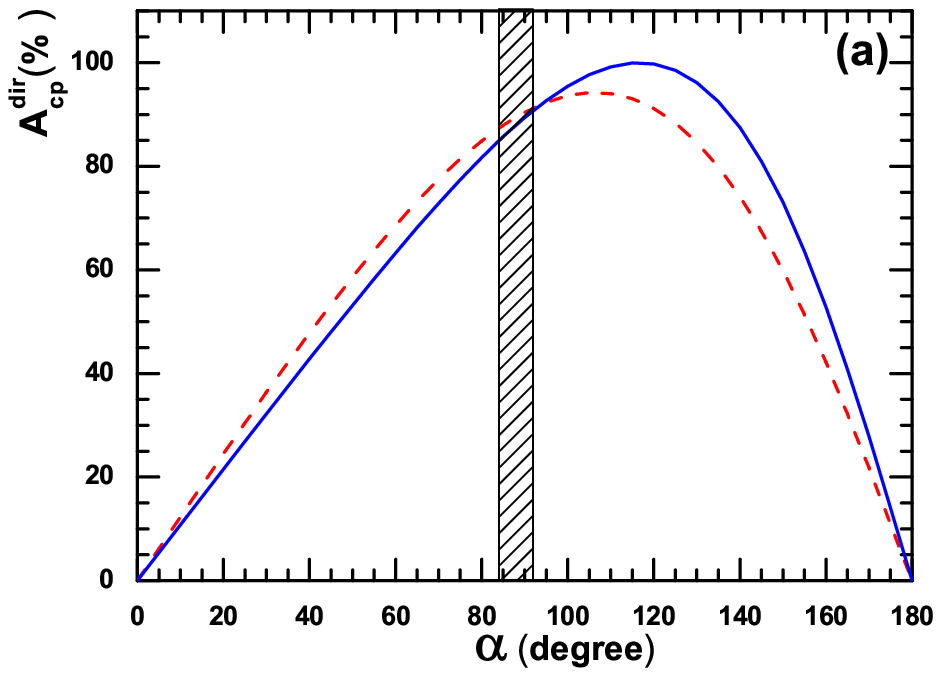}
\includegraphics[scale=0.6]{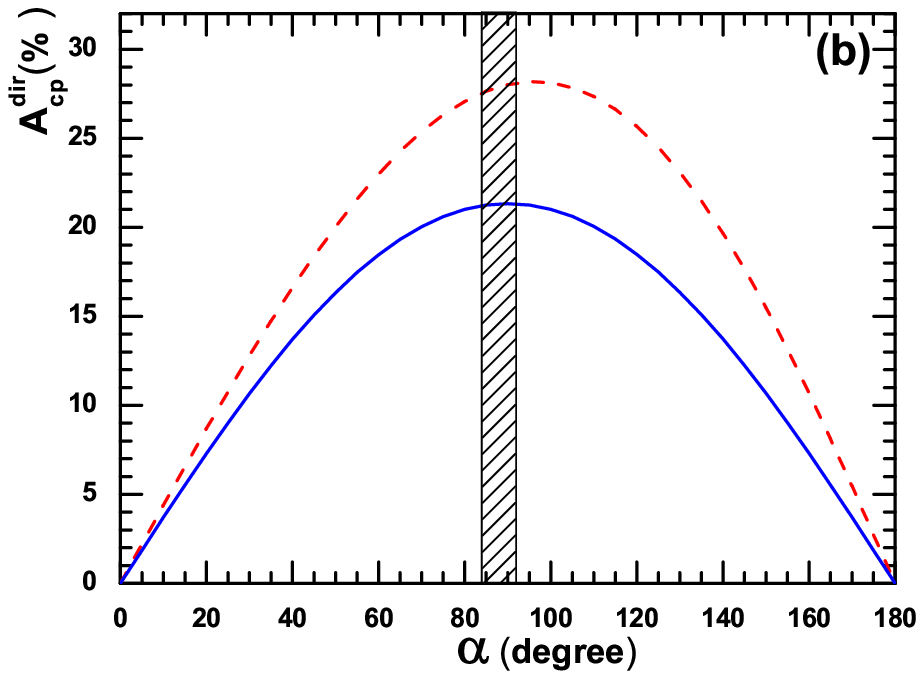}
\vspace{0.3cm} \caption{The dependence of the direct CP-violating asymmetries for $\bar B_s^0\to K^{*0}_0(1430)\pi^0$ (a)
and $\bar B_s^0\to K^{*+}_0(1430)\pi^-$ (b) on the
Cabibbo-Kobayashi-Maskawa angle $\alpha$. The dashed (solid) curves are plotted in scenario I (II).
The vertical bands show the range of $\alpha$: $89.0^{+4.4}_{-4.2}$ \cite{pdg10}.}\label{fig3}
\end{center}
\end{figure}

Using the calculated ratio $z$ and strong phase $\delta$, it is easy to calculate the numerical values
of $\acp^{dir}$ (in unit of $10^{-2}$) in two scenarios, which are the listed in Table 2.
The uncertainties are mainly from  the $B_s$ meson shape parameter $\omega_b=0.5\pm0.05$,
the Gegenbauer moments $B_1$ and $B_3$ of the scalar meson $K^*_0$, the CKM angle $\alpha=(100\pm20)^\circ$. For the decay mode $K^{*+}_0(1430)\pi^-$, as discussed above,
the value of
$(1+2z\cos\alpha\cos\delta+z^2)$ is close to $1$ in both scenarios, so the corresponding
direct CP-violating asymmetry is almost proportional to $\sin\alpha$, that is to say it attains the maximum near $90^\circ$ (shown in Fig.2b).
For the decay mode $K^{*0}_0(1430)\pi^0$, it receives a very large direct CP-violating asymmetry in both scenarios. It is not strange:
one can recall that the channel $\bar B^0_s\to K^{0}\pi^0$ also receives a large direct CP-violating asymmetry, $(59.4^{+7.9}_{-12.5})\%$ in the pQCD approach \cite{lucd0},
about $(41.6^{+47.1}_{-55.8})\%$ predicted by QCDF approach \cite{Beneke}.

\section{Conclusion}\label{summary}

In this paper, we calculate the branching ratios and the CP-violating
asymmetries of decays $\bar B_s^0\to K^*_0(1430)\pi$
in the PQCD factorization approach.
Using the decay constants and light-cone distribution amplitudes
derived from QCD sum-rule method, we find that
although these two decays are both
tree-dominated, the ratio of their penguin to tree contributions are very different, there is only a few percent for the decay
$\bar B_s^0\to K^{*+}_0(1430)\pi^-$, while about
$37\%$ in scenario I, even $51\%$ in scenario II for the decay $\bar B_s^0\to K^{*0}_0(1430)\pi^0$, which results
these two decays have very different values in the branching ratios and the direct CP asymmetries: The branching ratio
of the decay $\bar B_s^0\to K^{*+}_0(1430)\pi^-$ is at the order of $10^{-5}$, its direct CP asymmetry is about $(20-30)\%$. While for the decay
$\bar B_s^0\to K^{*0}_0(1430)\pi^0$,
its direct CP-violating asymmetry is very large and about $90\%$, but it is difficult to measure it, because the branching ratio for this channel is small and only $10^{-7}$
order. Between these two scenarios, there is less different for the results of the channel $\bar B_s^0\to K^{*0}_0(1430)\pi^0$, while larger different for those of
$\bar B_s^0\to K^{*+}_0(1430)\pi^-$, especially from the branching ratio. Because its branching ratio is proportional to
the modular square of the tree amplitude, which
for scenario II is about $3.1$ times that of scenario I. So the decay $\bar B_s^0\to K^{*+}_0(1430)\pi^-$ can be used to determine the inner
structure of $K^*_0(1430)$ by comparing the theoretical prediction with the future experimental data.

\acknowledgments
This work is partly supported by the National Natural Science Foundation of China under Grant No. 11047158,
and by Foundation of Henan University of Technology under Grant No. 2009BS038. The author would like to thank
Cai-Dian L\"u for helpful discussions.

\end{document}